\documentclass{elsart}
\usepackage{epsfig}
\usepackage{amssymb}

\begin{document}

\begin{frontmatter}

\title{Wigner function for discrete phase space: exorcising ghost images}

\author{Arturo Arg\"uelles}, \author{Thomas Dittrich}
\ead{tdittrich@unal.edu.co}

\address{Departamento de F{\'\i}sica,
  Universidad Nacional de Colombia,\\Santaf\'e de Bogot\'a, Colombia}

\begin{abstract}
We construct, using simple geometrical arguments, a Wigner function
defined on a discrete phase space of arbitrary integer Hilbert-space
dimension that is free of redundancies. ``Ghost images'' plaguing
other Wigner functions for discrete phase spaces are thus revealed as
artifacts. It allows to devise a corresponding phase-space propagator
in an unambiguous manner. We apply our definitions to eigenstates
and propagator of the quantum baker map. Scars on unstable periodic
points of the corresponding classical map become visible with
unprecedented resolution. 

\end{abstract}

\end{frontmatter}

\section{Introduction}
\label{sec1}
Since the conception of the Wigner function in 1932 \cite{Wig}, phase-space
representations of quantum mechanics have found a wide range of
applications. Notably in complex quantum dynamics, they have become
popular as they allow to compare classical and quantum dynamics on the
same footing. The Wigner function in particular has become the
standard tool whenever the full information contained in the density
matrix, including coherences, has to be represented. 

In many cases the ``quantum phase space'' underlying the Wigner
function is fundamentally different from the corresponding classical
one in that it is \emph{discrete}. This occurs, for example, if both
classical phase-space variables are cyclic, i.e., if the classical
phase space has the topology of a torus so that the periodicity of
either variable implies quantization of the other. Likewise, it is
inevitable once, e.g., in numerical implementations of wave-packet
scattering, the spatial variable is reduced to a discrete lattice
and at the same time restricted to some finite total length, with
periodic boundary conditions. Finally, discrete phase spaces allow to
define a perfectly self-consistent quantum mechanics living on a
finite-dimensional Hilbert space \cite{GaRu}.

A straightforward generalization of the original definition to
the discrete case leads to a Wigner function that contains the
information of the density operator in several replicas. This
redundancy is reflected in so-called ghost images of features standing
out in the quantum state \cite{Kol,MiPaSa}. 
Modified definitions that avoid these redundancies have been proposed
in \cite{Woo,AgBr}. We here start from the version in
Ref.~\cite{AgBr}, giving a transparent derivation in simple geometric
terms that readily generalizes to \emph{arbitrary} (not necessarily
odd or even prime) integer Hilbert-space dimensions. It allows to
construct a propagator on the discrete phase space, which can be
defined unambiguously only for a Wigner function free of
redundancy. As an illustrative example, we apply it to the quantum
baker map \cite{BaVo}. A more detailed account of our findings will be
published elsewhere \cite{ArDi}. 

\section{Non-redundant Wigner function and propagator}
\label{sec2}
The original definition of the Wigner function \cite{Wig}, $W(p,q) =
h^{-1}\int_\infty^\infty {\rm d}q'\,$ ${\rm e}^{-{\rm i}pq'/\hbar}$
$\rho(q+q'/2,q-q'/2)$ for a density operator $\hat\rho$ consists of
two steps, a transformation to sum-and-difference coordinates $q =
(q_1+q_2)/2$, $q' = q_1-q_2$ of the density matrix $\rho(q_1,q_2) =
\langle q_1|\hat\rho|q_2\rangle$, followed by a Fourier transform
along $q'$. The difficulty in applying this to a discrete spatial
basis $|n\rangle$, $n = 0,\ldots,N-1$ arises from the fact that in
$(q,q')$-space the original square lattice appears as ``base-centered
cubic'', with even and odd points in terms of $n_1\pm n_2$. Augmenting
it to become ``simple cubic'' again by introducing fictitious extra
lattice points where $\rho(n,n') = 0$ lead, upon Fourier
transformation (of size $2N$), to a fourfold redundancy of the
resultant Wigner function \cite{Kol,MiPaSa}, call it $W_{\rm
double}({\bf m})$, abbreviating the discrete phase-space vector ${\bf
m} = (\lambda,n)$. We remove the redundancy as follows: By an
additional Fourier transform (size $2N$), now in the $q$-direction,
the redundancy takes the explicit form of an identical repetition, up
to sign changes, of the coefficients in four $N\times N$-blocks within
the total $2N\times 2N$ ``Brillouin zone''. We cut out a single
$N\times N$-block, chosen centered around the origin so as to preserve
twofold symmetries that reflect the reality of the Wigner function, by
multiplying with a two-dimensional discrete box function. Upon undoing
the additional second Fourier transform (now of size $N$ only), the
Wigner function is recovered, \emph{free of redundancies}.

Cutting out the $N\times N$-block from the
Fourier-transformed Wigner function $\widetilde W_{\rm double}$ is
equivalent to a convolution of $W_{\rm double}$ with a
Fourier-transformed box function $\tilde S$ 
\begin{eqnarray}\label{wigboxfunc}
W({\bf m}) &=& \sum_{\lambda',n'=-N}^{N-1}
W_{\rm double}(2{\bf m}-{\bf m}') \tilde S({\bf m}'),\\
\tilde S({\bf m}) &=& \tilde s_\lambda \tilde s_n, \quad
\tilde s_k = \left\{\begin{array}{l@{\quad}l}
\delta_{k\,{\rm mod}\,2N} &k\mbox{ even},\\
{\sin(\pi k/2)\over N\sin(\pi k/2N)}
{\rm e}^{-{\rm i}\pi k/2N}&k\mbox{ odd}.
\end{array}\right.
\end{eqnarray}

Apart from revealing ``ghost images'' as artifacts and removing them,
this procedure becomes indispensable once a propagator of the Wigner
function is to be defined in an unambiguous manner. The reason is that
with a redundant Wigner function, there is no unique way of relating
the coefficients of the final Wigner function to those of the initial
one. Denoting the transformation outlined above symbolically as
$T_{\rm W}$, so that $W({\bf m}) = T_{\rm W} \rho({\bf n})$, with
${\bf n} = (n_1,n_2)$, we find for the discrete Wigner function
propagator 
\begin{equation}\label{wigrhoprop}
K_{\rm W}({\bf m}'',t'';{\bf m}',t') = 
T_{\rm W} K({\bf n}'',t'';{\bf n}',t') T_{\rm W}^{-1},
\end{equation}
where $K({\bf n}'',t'';{\bf n}',t')$ propagates the density matrix
from time $t'$ to $t''$, i.e., $\rho({\bf n}'',t'') = \sum_{\bf n}'
K({\bf n}'',t'';{\bf n}',t') \rho({\bf n}',t')$.

If the time evolution is unitary, generated by a Hamiltonian $\hat H$,
the propagator can be expressed in terms of energy eigenstates $\hat H
|\alpha\rangle = E_\alpha |\alpha\rangle$ as $K({\bf n}'',t'';{\bf
n}',t') = \sum_{\alpha,\beta} {\rm e}^{-{\rm i}(E_\alpha-E_\beta)
(t''-t')/\hbar} \rho_{\alpha,\beta}^*({\bf n}'')
\rho_{\alpha,\beta}({\bf n}')$, denoting $\rho_{\alpha,\beta}({\bf n})
= \langle\alpha|n_1\rangle\langle n_2|\beta\rangle$. Defining Wigner
eigenfunctions as $W_{\alpha,\beta}({\bf m}) = T_{\rm W}
\rho_{\alpha,\beta}({\bf n})$, we have for the Wigner propagator
\begin{equation}\label{wigpropeifunc}
K_{\rm W}({\bf m}'',t'';{\bf m}',t') = N^{-1}\sum_{\alpha,\beta}
{\rm e}^{-{\rm i}(E_\alpha-E_\beta)(t''-t')/\hbar} 
W_{\alpha,\beta}^*({\bf m}'') W_{\alpha,\beta}({\bf m}').
\end{equation}
Equation (\ref{wigpropeifunc}) allows for an easy direct numerical
access to the propagator once enough eigenfunctions, e.g., in position
representation, are known.

\section{Wigner eigenfunctions and propagator for the quantum baker
map} 
\label{sec3}
\begin{figure}[t]
 \centerline{
  \psfig{figure=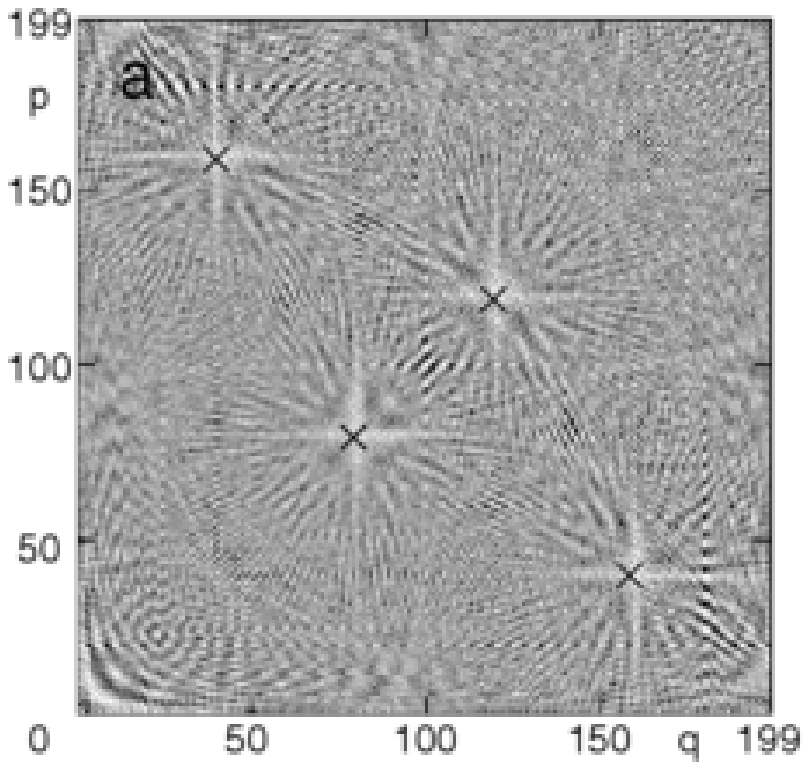,width=65mm}\hspace{0mm}
  \psfig{figure=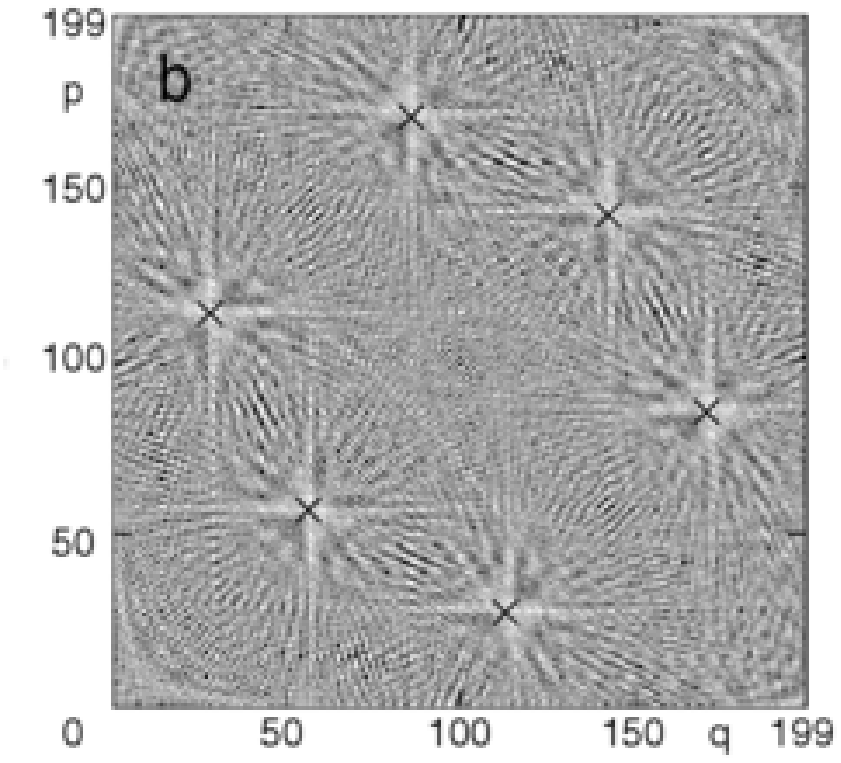,width=65mm}\hspace{0mm}
 }
\vspace*{1mm}
 \centerline{
  \psfig{figure=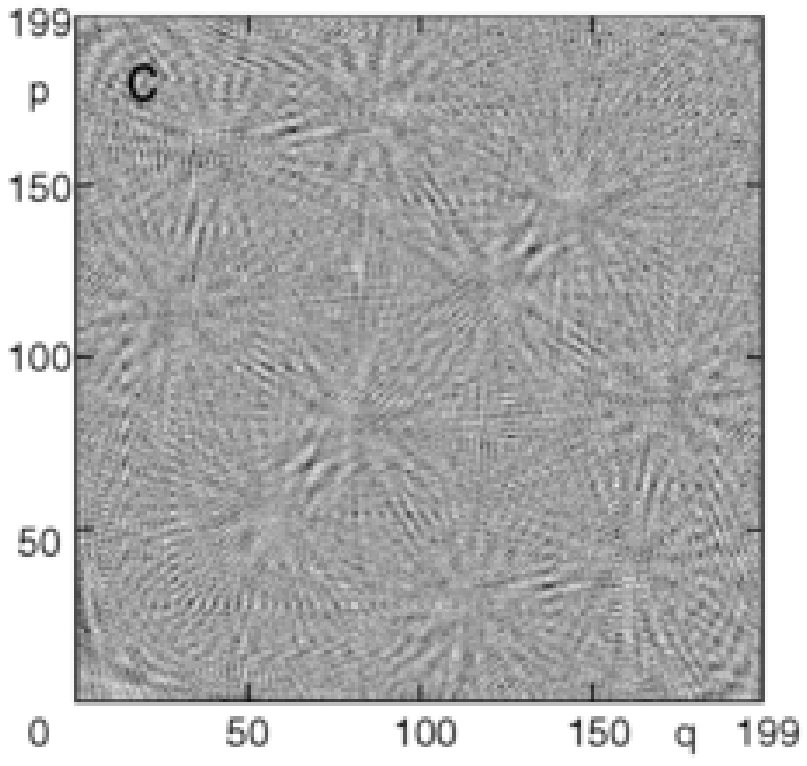,width=65mm}\hspace{0mm}
  \psfig{figure=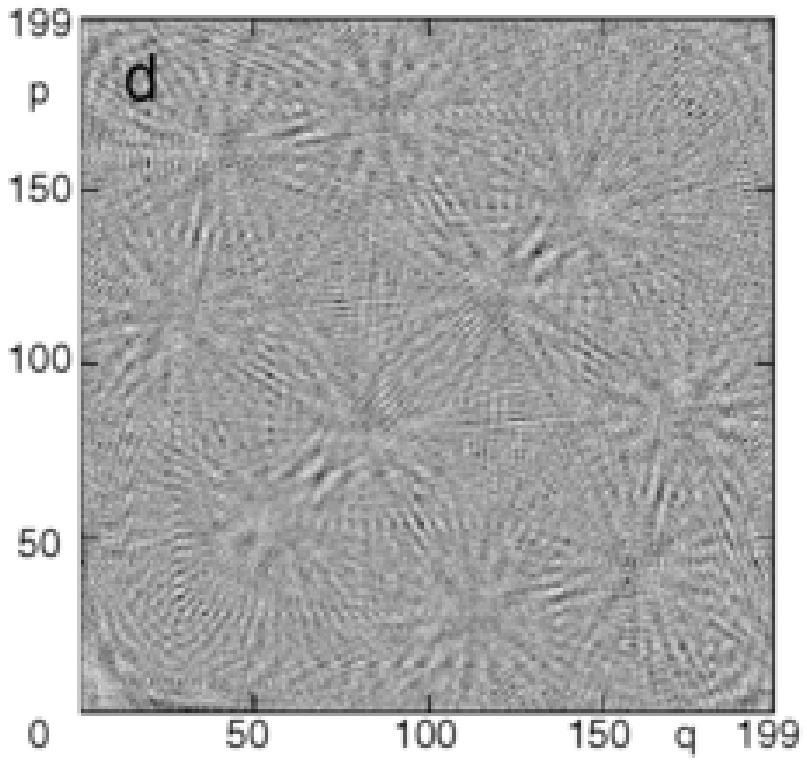,width=65mm}\hspace{0mm}
 }
 \caption{\label{qbmeigenst}
Wigner functions according to Eq.~(\protect\ref{wigboxfunc}) for
eigenstates of the quantum baker map,
Eq.~(\protect\ref{qubaker}). Panels a,b show diagonal Wigner
eigenfunctions $W_{\alpha,\alpha}$ for (a) $\alpha = 88$ and (b)
$159$. Panels c,d are real and imaginary parts, respectively, of the
off-diagonal Wigner eigenfunction $W_{88,129}$. Hilbert-space dimension
is $N = 200$. Greyscale ranges from negative (dark) to positive
(bright) values. Unstable periodic orbits of the corresponding
classical map underlying the scars have been marked with crosses.
}
\end{figure}

\begin{figure}[t]
 \centerline{
  \psfig{figure=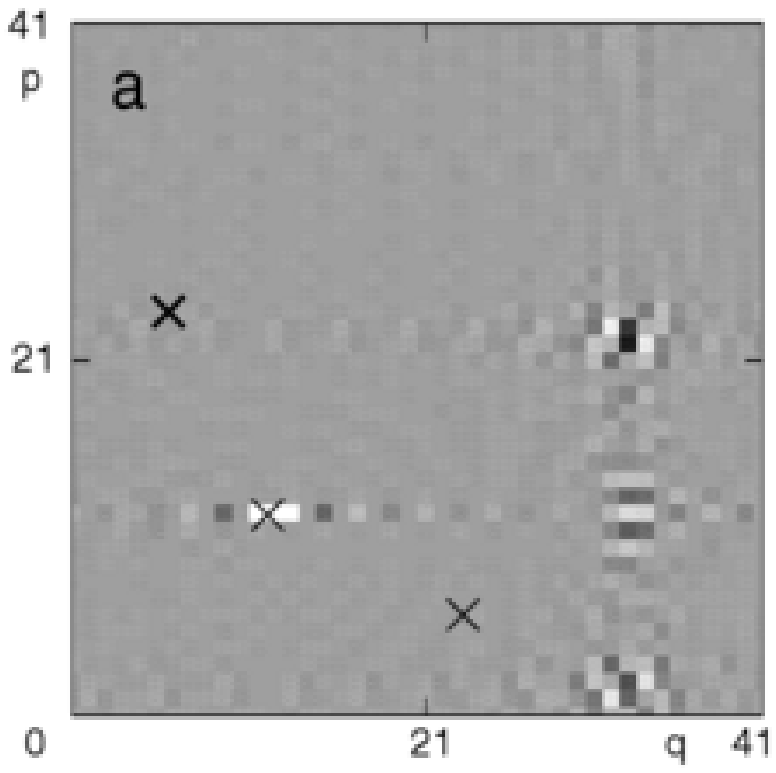,width=65mm}\hspace{0mm}
  \psfig{figure=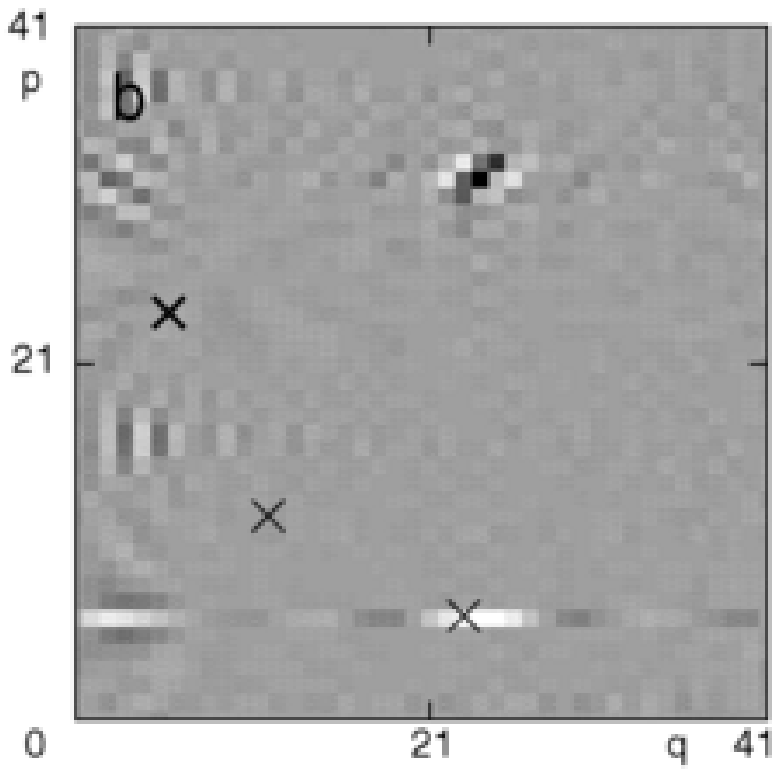,width=65mm}\hspace{0mm}
 }
 \caption{\label{qbmprop}
Propagator of the Wigner function according to
Eq.~(\protect\ref{wigpropeifunc}) for initial condition ${\bf m}' =
(6,25)$ (bold cross) on the period-3 unstable periodic orbit as marked
in Fig.~\protect\ref{qbmeigenst}, after (a) $1$ and (b) $2$ applications
of the same quantum baker map as in
Fig.~\protect\ref{qbmeigenst}. Hilbert-space dimension is $N =
42$. Other parameters and greyscale as in Fig.~\protect\ref{qbmeigenst}.
}
\end{figure}

The baker transformation \cite{BaVo} is arguably the simplest example
of a fully chaotic map of the unit square to itself. A quantum version
has been suggested in Ref.~\cite{BaVo} in terms of Fourier
transformations $F_M$, $(F_M)_{m,n} = M^{-1/2} {\rm e}^{-2\pi{\rm
i}mn/M}$ of sizes $M = N$ and $N/2$, where $N$ (even) denotes the
Hilbert-space dimension, as follows, 
\begin{equation}\label{qubaker}
B_N = F_N^{-1} 
\left(\begin{array}{cc}F_{N/2}&0\\ 0&F_{N/2}\end{array}\right),
\end{equation}
so that wavefunctions map $\psi'(n') = \sum_{n=0}^{N-1}
(B_N)_{n'n}\psi(n)$, and the density operator accordingly. We have
calculated diagonal Wigner eigenstates $W_{\alpha,\alpha}$ and
off-diagonal ones $W_{\alpha,\beta}$ and used them to construct the
Wigner propagator equivalent to the unitary map (\ref{qubaker})
according to Eq.~(\ref{wigpropeifunc}). The eigenstates $\alpha = 88$
(Fig.~\ref{qbmeigenst}a) and $129$ (\ref{qbmeigenst}b) clearly show
scars of classical unstable periodic orbits (crosses) which up to now
could be identified only in the corresponding Husimi functions
\cite{Sar}, with considerably inferior resolution. Figures
\ref{qbmeigenst} c,d show, respectively, the real and imaginary part
of the off-diagonal eigenstate $W_{88,129}$, scarred simultaneously by
both orbits that appear in the eigenstates $W_{88,88}$ and
$W_{129,129}$. 

The Wigner propagator is depicted in Fig.~\ref{qbmprop}, for fixed
initial ${\bf m}' = (6,25)$ at $t' = 0$, the leftmost point of one of
the period-3 orbits scarring $W_{129,129}$ (Fig.~\ref{qbmeigenst}b),
at final times $t'' = 1$ (panel a) and $t'' = 2$ (b). The strong
localization of the propagator on the underlying classical orbit is
evident, as are quantum coherence effects reflected in oscillatory
patterns far off the classical orbit.

\section{Conclusion}
\label{sec4}
The redundance-free Wigner function for discrete phase spaces proposed
here enables to identify classical and other phase-space structures in
states and time evolution of quantum systems with finite-dimensional
Hilbert space, without being marred by artificial ghost images. It
facilitates developping and assessing semiclassical approximations for
the propagator in such systems \cite{DiSaVi}, providing an alternative
point of view for the analysis of coherent structures like ``quantum
carpets'' \cite{Ber,GrRoSc}. 

The method allows to represent signals simultaneously in time and in
frequency space, in all applications where data are both discrete and
periodic in time (or space) such as in optics, acoustics, or other
fields possibly far away from quantum mechanics.

\section{Acknowledgements}
\label{sec5}
We benefitted from discussions with Andrey Kolovsky, J\"urgen Korsch,
and Bilha Segev. Financial support from the Volkswagen Foundation
(contract I/78235) is gratefully acknowledged.

\end{document}